\documentclass[conference,usenatbib]{basi}
\usepackage[T1]{fontenc}
\usepackage[british]{babel}
\usepackage[varg]{txfonts}
\usepackage{rotating}
\usepackage{dcolumn}
\begin{document}
\title[Spectro-temporal evolution of 2011 outburst of IGR J17091-3624]{Spectro-temporal evolution during the 
	onset-phase of the 2011 outburst of IGR J17091-3624 -- Implications on accretion disk dynamics}
\author[N.~Iyer \& ~A.~Nandi]%
{Nirmal Iyer$^{1,2}$\thanks{email: \texttt{nirmal@physics.iisc.ernet.in}},
	          A. Nandi$^{1}$\\
			         $^1$Space Astronomy Group, SSIF, ISITE Campus, ISAC, Bangalore, India\\
					        $^2$Dept. of Physics, Indian Institute of Science, Bangalore, India}

							\pubyear{2013}
							\volume{**}
							\pagerange{**--**}

							\date{Received --- ; accepted ---}

							\maketitle
							\label{firstpage}

							\begin{abstract}
							  We re-analysed the archival data of RXTE / INTEGRAL / Swift satellites at the onset 
							  phase of the 2011 outburst of the X-ray source IGR J17091-3624. The evolution of the
							  spectral and temporal properties of the source in this phase clearly exhibits state
							  transition as Hard (HS) $\rightarrow$ Hard-Intermediate (HIMS) $\rightarrow$ Soft-Intermediate (SIMS) $\rightarrow$ 
							  Soft (SS) state, before entering the variability phase (VP).
							  We attempt to understand the evolution of X-ray features and the state transitions 
							  based on two different types of accreting material (i.e, the Keplerian and sub-Keplerian flow).
							\end{abstract}

							\begin{keywords}
							  accretion disks, accretion physics, X-ray sources, black holes  
							\end{keywords}

							   \section{Introduction}\label{s:intro}

							   The X-ray source IGR J17091-3624 was first observed in 2003 by IBIS on board the
							   INTEGRAL satellite \citep{2003ATel..149....1K}. In early 2011, the source brightened
							   again \citep{2011ATel.3144....1K}. The subsequent observations (in 2011) have led to
							   the discovery of similar intensity-time variations (i.e. the variability phase) in this source as shown by 
							   GRS 1915+105. We make an attempt to re-look at the initial few days ($\sim$ 40 days)
							   of the outburst in 2011. We believe that the onset phase of the outburst though 
							   looked at by others (\citet{2011arXiv1105.4694P} and \citet{2012MNRAS.422.3130C}),
							   can be looked at in greater detail to understand the disk dynamics. In the following
							   sections, we outline the observations and analysis techniques and draw inferences 
							   from the results based on our analysis.

							   \begin{figure}
								 \centering{\includegraphics[width=10cm]{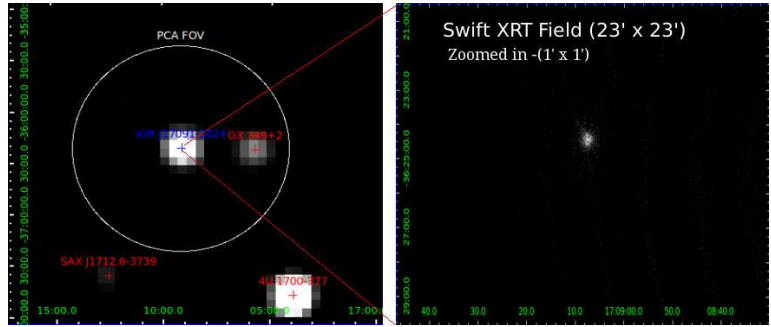} }

								 \caption{IGR J17091-3624 and surrounding field (left panel INTEGRAL IBIS, 
								   right panel Swift XRT). As seen, the PCA FOV has two sources leading 
								   to source confusion. \label{f:image}}
							   \end{figure}

							   \section{Observation and Analysis}\label{s:Obs}
							   \vspace{-5mm}
							   The observations we used in the rising phase of the outburst were from the dates 
							   \texttt{2011 Feb 03} to \texttt{Mar 14} ($\sim$ 40 days). In this duration, we used 
							   the archival data from Swift, INTEGRAL and RXTE satellites. The source field being 
							   rather crowded (Fig. \ref{f:image}; see also \citet{2011arXiv1105.4694P}), the RXTE 
							   PCU observations were contaminated until \texttt{Feb 23} by the nearby bright X-ray 
							   LMXB source GX 349+2. Hence only the central frequency of the QPO was extracted from 
							   the contaminated RXTE data. The QPO feature of the PDS was confirmed to be from the 
							   IGR J17091 source by looking at the absence of such a feature from simultaneous 
							   observation of the nearby source (GX 349+2) using INTEGRAL JEM-X for the same energy band.

							   \vspace{-3mm}
							   Temporal analysis was done using Swift XRT (0.5 -- 10 keV) and RXTE data sets with 
							   the above mentioned constraints. For analysing the data, \textbf{GHATS v1.1} \footnotemark
							   , a customised IDL based timing package was used. The Swift XRT data extraction was 
							   done using \textbf{HEASOFT 6.12} and the ftools package \textbf{xrtpipeline}. The 
							   INTEGRAL data reduction was done using the \textbf{OSA 10.0} software. We used 
							   \textbf{XSPEC v12.7.1} for spectral fitting. For 0.5 -- 10 keV Swift XRT data, we used 
							   \texttt{diskbb} and \texttt{powerlaw} modified by an absorption column (\texttt{phabs}). 
							   For simultaneous Swift and INTEGRAL data (0.5 -- 150 keV), the \texttt{powerlaw} model 
							   was replaced by the \texttt{cutoffpl} model, which enabled us to study the cutoff/fold 
							   energy evolution with time. The data in the rising phase (until the observation of 
							   \texttt{Feb 22}) did not require additional \texttt{diskbb} component, and fit well 
							   with a \texttt{powerlaw} (or \texttt{cutoffpl}) and a \texttt{phabs} model. For modelling the power spectra 
							   as well as the QPO frequencies, a combination of the \texttt{lorentz} models were used.

							   \footnotetext[1]{ http://www.brera.inaf.it/utenti/belloni/GHATS\_Package/Home.html }

							   \vspace{-3mm}
							   \section{Results}\label{s:results}
							   \vspace{-5mm}
							  
							   \begin{figure}
								  \begin{minipage}[t]{6cm}
								  \begin{center}
								  	  \includegraphics[width=6cm]{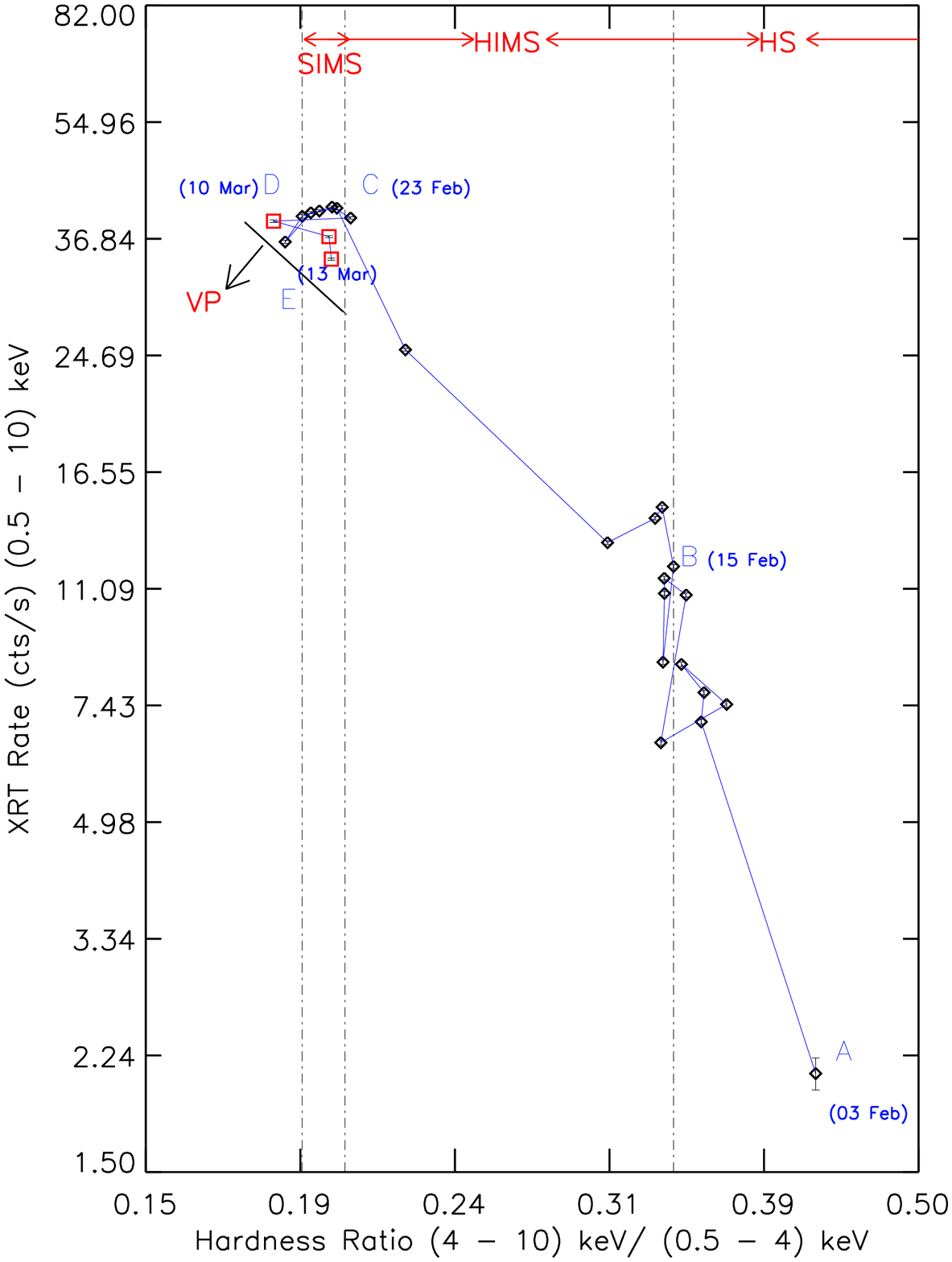}
									  \end{center}
								  \end{minipage}
								  \begin{minipage}[t]{5.5cm}
								  \begin{center}
									\vspace{-3.14in}
								      \includegraphics[width=5cm,height=3.81cm]{nirmaliyer_02b.ps} \\
									  \includegraphics[width=5cm,height=3.81cm]{nirmaliyer_02c.ps}
									  \end{center}
								  \end{minipage}
								  \vspace{-5mm}
								  \caption{In Hardness Intensity Diagram (left), the states are demarcated by the 
									vertical lines, except the Soft state (marked by red boxes) and the VP.
									Energy dependence of QPO (right) -- the QPO is visible in the 
									(4 -- 30) keV range and is not seen in the soft photons (0.5 -- 4) keV, which 
								    mostly arise from the Keplerian accretion disk. Power spectra are from 23 Feb (PCU)
								    and 24 Feb (XRT) observations.} 
								  \label{fig:hid}
								  \vspace{-3mm}
							  \end{figure}

\begin{figure}
								  \begin{minipage}[t]{6cm}
								  	  \includegraphics[width=6cm]{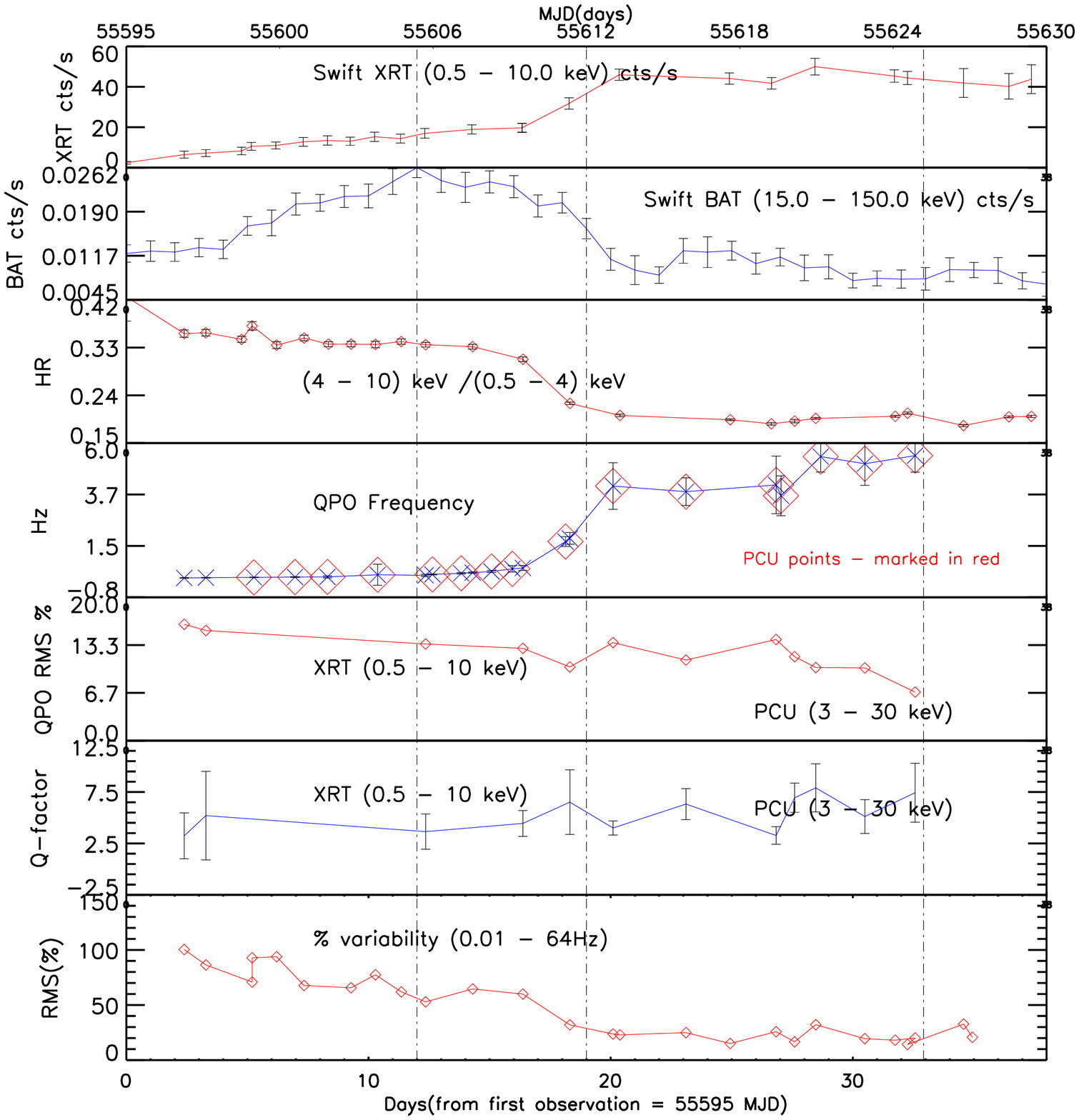}
								  \end{minipage}
								  \begin{minipage}[t]{6cm}
								      \includegraphics[width=6cm]{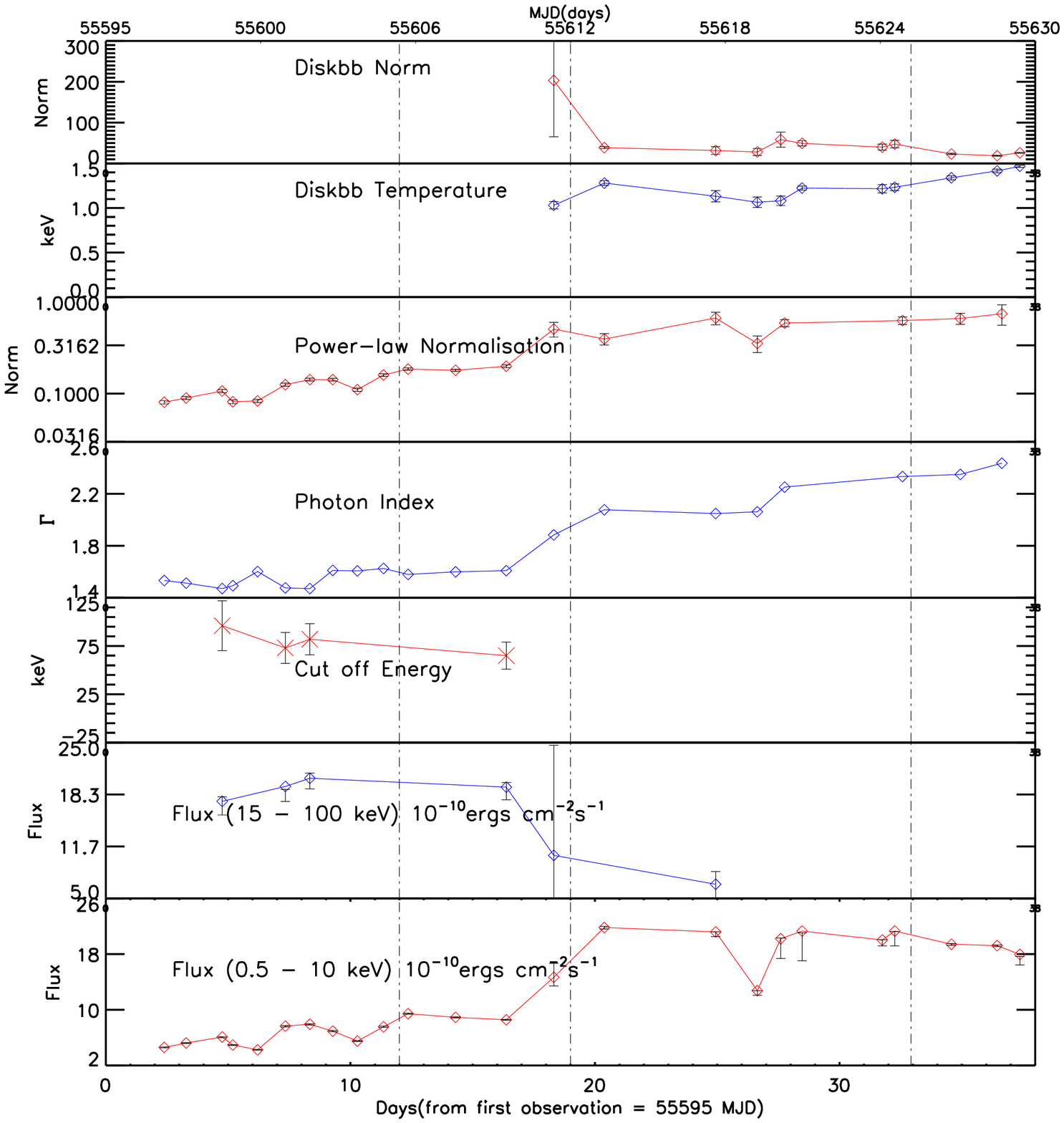}
								  \end{minipage}
								\caption{Evolution of temporal (left) and spectral (right) features of the 2011 outburst of 
								 IGR J17091-3624
								  (except the variability phase). 
								  The vertical lines indicate the state transitions observed during the outburst (See Fig. \ref{fig:hid} 
								   and text for details).}
								\label{fig:variation}
							  \end{figure}

\begin{figure}[b!]
								\centering
								\includegraphics[height=7cm,angle=270]{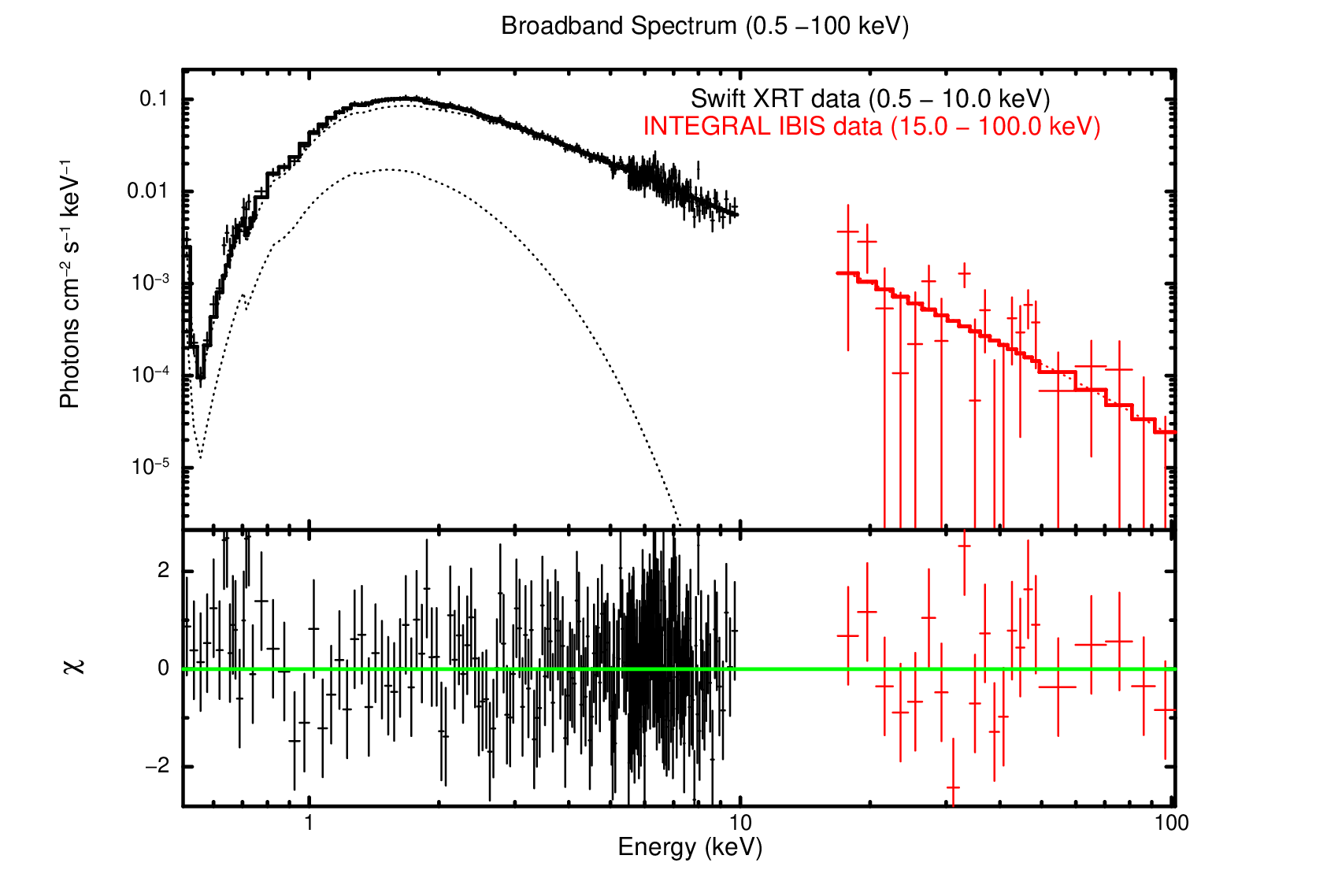}
								\caption{Broadband Spectrum of \texttt{2011 Feb 22} observation fitted with 
								  \texttt{phabs*(diskbb + cutoffpl)} model.}
								\label{fig:bbspec}
					          \end{figure}

							   In Fig. \ref{fig:hid}, we have plotted the Hardness Intensity Diagram (HID) that was 
							   obtained by plotting the XRT count rate (0.5 -- 10 keV) against the ratio of counts 
							   in the 4 -- 10 keV band to that in the 0.5 -- 4 keV band. The evolution of HID is 
							   clearly seen to vary with distinct transition between states. The transitions are also
							   seen in the variations of hardness, QPO frequency, soft flux, photon index ($\Gamma$) and 
							   overall RMS as shown in Fig. \ref{fig:variation}. Hence, we propose that the evolution is 
							   along the lines as seen in other black hole binaries, as outlined -- \\\textbf{(a) Hard state}
							   (from \texttt{03 Feb -- MJD 55595.90}) -- marked by mostly constant low frequency QPO
							   ($\sim$ 0.1Hz) and $\Gamma$ ($\sim$ 1.6) with high values of HR ($\sim$ 0.35). \\
							   \textbf{(b) Hard Intermediate state} (from \texttt{15 Feb -- MJD 55607.25}) -- marked by decrease 
							   from the peak value in the BAT counts, gradual increase in the QPO frequency, 
							   $\Gamma$ and a significant drop in the HR value indicative of a state 
							   transition. \\\textbf{(c) Soft Intermediate state} (from \texttt{23 Feb -- MJD 55614.21}) -- 
							   marked by higher and very slowly increasing values of ($\Gamma$) 
							   and QPO frequency as compared to the previous states and the appearance of a 
							   disk component, as shown in Fig. \ref{fig:bbspec} . \\\textbf{(d) Soft State} (\texttt{10 Mar to 13 Mar}) -- 
							   marked by absence of QPO (0.5 -- 30 keV) and soft spectra with higher $\Gamma$ and disk temperature, 
							   as shown in Fig. \ref{fig:variation}
							   \\ It is also observed that the QPO is 
							   not seen in the soft photons (0.5 - 4 keV), once the disk is observed in the spectrum 
							   (right panel of Fig. \ref{fig:hid}). All these variations (temporal as well as spectral) 
							   can be interpreted in terms of the Two Component Advective Flow (TCAF) model proposed by 
							   \citet{1995ApJ...455..623C}. The right-most state in the HID corresponding to the 
							   HS is mostly dominated by the Sub-Keplerian flow with hot electron 
							   plasma causing the hard emission. The central states in the HID -- the HIMS 
							   (with increasing Keplerian component in the flow) and the SIMS (with 
							   comparable Keplerian and Sub-Keplerian inflow matter) states have softer emission with 
							   higher fluxes, and are followed by a disk dominated Soft state. After \texttt{13 Mar}, the 
							   system exhibits a variety of X-ray oscillatory features ($\rho, \alpha, \nu,\mu$ and other such 
							   classes; see also \citet{2011ApJ...742L..17A}) like GRS 1915+105 but seems to be confined (as 
							   indicated by the arrow; see also \citet{2012MNRAS.422.3130C}) in 
							   the Soft Intermediate state of the declining phase of the outburst (marked as VP in HID).

							  The spectrum of the source, as obtained by simultaneously fitting of INTEGRAL and Swift 
							  spectra is shown in Fig. \ref{fig:bbspec}. We have attempted to fit the broad-band 
							  (0.5 -- 100 keV) spectrum using a phenomenological model consisting of a multi-color 
							  disk and power-law with an exponential roll-off (or cutoffpl) modified by columnar absorption. 
							  Similar kind of fitting of the spectrum using the TCAF model and its implications on the disk 
							  dynamics and system parameters are investigated further in \citealt{2013Fut..test} (in prep.).

					   \section{Conclusions}\label{s:concls}
					   \vspace{-5mm}
					   The movement of the system through the HID in distinct states can be interpreted in terms of a 
					   sub-Keplerian hot flow responsible for the hard power-law component of the spectrum, and a Keplerian 
					   flow giving the soft photons as per the classical multi-color disk model. The QPO variations can 
					   be modelled using the movement of an oscillatory shock front in the TCAF model (see 
					   \citealt{2013Fut..test}). The continuous inward drift of this shock front causes the 
					   QPO frequency to vary as seen in Fig. \ref{fig:variation}, till the system transits to the SIMS. Here,
					   the QPO frequency initially stabilises indicating that the shock front does not move in further, and then
					   dies out, indicating a Keplerian disk dominated Soft state.
					   The variations in the spectral features can also be explained under the TCAF paradigm as
					   already attempted for the case of GX 339-4 \citep{2012A&A...542A..56N}. The soft photons from the 
					   disk, are not seen in either the HS or the HIMS indicating negligible inflow 
					   from the Keplerian component in these states. The modelling of the spectro-temporal 
					   correlations (the QPO evolution and $\Gamma$-QPO relation) and the 
					   broad-band spectrum gives preliminary indication that the source is a massive ($>$ 10M$_{\odot}$) 
					   black hole \citep{2013Fut..test}, like the galactic black hole source GRS 1915+105. Based 
					   on this similarity in mass and temporal variations, we propose that the source GRS 1915+105 
					   could also be currently locked in the variability phase (VP) of its evolution, and that some 
					   time in the past had undergone evolution from the Hard to the Intermediate to the 
					   present state (i.e., variability phase).

					\section*{Acknowledgments}
					  The authors would like to thank Dr. S. Seetha (Space Science Office, ISRO HQ) and
					  Dr. P. Sreekumar (Space Astronomy Group, ISAC) for helping them make use of the facilities 
					  at ISAC to do this research and to be able to attend the conference.

\end{document}